\documentclass[letterpaper,twocolumn,10pt]{article}
\usepackage{usenix-2020-09}

\usepackage{amsmath}
\usepackage{booktabs}
\usepackage{subcaption}
\usepackage{xspace}
\usepackage{xcolor}
\usepackage{listings}
\usepackage[textsize=scriptsize]{todonotes}
\usepackage{tikz}
\usepackage{forloop}
\usepackage{xifthen}
\usepackage{multirow}
\usepackage{mathtools}
\usepackage{amssymb}
\usepackage{amsmath}
\usepackage{enumitem}
\usepackage{balance}
\usepackage{circledsteps}
\usepackage{xurl}
\usepackage{hyperref}

\definecolor{codegreen}{rgb}{0,0.6,0}
\definecolor{codegray}{rgb}{0.5,0.5,0.5}
\definecolor{codepurple}{rgb}{0.58,0,0.82}
\definecolor{backcolour}{rgb}{0.95,0.95,0.92}

\lstdefinestyle{myStyle}{
    belowcaptionskip=1\baselineskip,
    breaklines=true,
    frame=none,
    backgroundcolor=\color{backcolour},   
    commentstyle=\color{codegreen},
    keywordstyle=\color{magenta},
    numberstyle=\tiny\color{codegray},
    stringstyle=\color{codepurple},
    basicstyle=\ttfamily\footnotesize,
    breakatwhitespace=false,         
    breaklines=true,                 
    keepspaces=true,                 
    numbers=left,       
    numbersep=5pt,                  
    showspaces=false,                
    showstringspaces=false,
    showtabs=false,                  
    tabsize=2,
}

\begin{document}

\date{}

\title{
    \Large \bf Tensor Memory Engine: 
    On-the-fly Data Reorganization for Ideal Locality
}

\author{
    {\rm Denis Hoornaert}\\
    Technical University of Munich\\
    denis.hoornaert@tum.de
    \and
    {\rm Cole Strickler}\\
    University of Kansas\\
    colestrickler@ku.edu
    \and
    {\rm Manos Athanassoulis}\\
    Boston University\\
    mathan@bu.edu
    \and
    {\rm Marco Caccamo}\\
    Technical University of Munich\\
    mcaccamo@tum.de
    \and
    {\rm Heechul Yun}\\
    University of Kansas\\
    heechul.yun@ku.edu
    \and
    {\rm Renato Mancuso}\\
    Boston University\\
    rmancuso@bu.edu
} 

\newcommand{\unamelong}{Tensor Memory Engine\xspace}
\newcommand{\uname}{TME\xspace}

\newcommand{\fig}[2][]{\hyperref[#2]{Fig.~\ref*{#2}#1}\xspace}

\newcommand{\tensor}{\ensuremath{T}}
\newcommand{\dimensions}{\ensuremath{N}\xspace}
\newcommand{\addrset}{\ensuremath{\mathbb{A}}\xspace}
\newcommand{\configset}{\ensuremath{\mathbb{C}}\xspace}
\newcommand{\conflength}{\ensuremath{l}\xspace}
\newcommand{\confstart}{\ensuremath{s}\xspace}
\newcommand{\confsize}{\ensuremath{w}\xspace}
\newcommand{\confcond}{\ensuremath{c}\xspace}
\newcommand{\confwidth}{\ensuremath{W}\xspace}
\newcommand{\confoffset}{\ensuremath{O}\xspace}
\newcommand{\boolset}{\ensuremath{\{\text{false}, \text{true}\}}\xspace}
\newcommand{\coordset}{\ensuremath{(\mathbb{R}_{\ge 0})^{\dimensions}}\xspace}

\newcommand{\corenum}{\ensuremath{n}}
\newcommand{\core}[1]{\ensuremath{C_{#1}}}
\newcommand{\coreset}{\ensuremath{\core{0}, \ldots, \core{\corenum}}}

\newcommand{\llc}{LLC\xspace}
\newcommand{\llclong}{Last-level Cache\xspace}
\newcommand{\fpga}{FPGA\xspace}
\newcommand{\fpgalong}{Field Programmable Gate Array\xspace}
\newcommand{\dram}{DRAM\xspace}
\newcommand{\dramlong}{Direct Random Access Memory\xspace}
\newcommand{\cci}{CCI\xspace}
\newcommand{\ccilong}{Cache Coherent Interconnect\xspace}
\newcommand{\kria}{Kria KV260\xspace}
\newcommand{\ultrascale}{UltraScale+\xspace}
\newcommand{\amd}{AMD-Xilinx\xspace}

\newcommand{\ci}{\emph{(i)}\xspace}
\newcommand{\cii}{\emph{(ii)}\xspace}
\newcommand{\ciii}{\emph{(iii)}\xspace}
\newcommand{\civ}{\emph{(iv)}\xspace}
\newcommand{\cv}{\emph{(v)}\xspace}
\newcommand{\cvi}{\emph{(vi)}\xspace}
\newcommand{\cvii}{\emph{(vii)}\xspace}
\newcommand{\cviii}{\emph{(viii)}\xspace}
\newcommand{\cix}{\emph{(ix)}\xspace}
\newcommand{\cx}{\emph{(x)}\xspace}

\newcommand{\eg}{e.g.,\xspace}
\newcommand{\ie}{i.e.,\xspace}
\newcommand{\wrt}{w.r.t.\xspace}
\newcommand{\tm}{\textsuperscript{\texttrademark}\xspace}

\newcommand{\rob}{ROB\xspace}
\newcommand{\roblong}{Re-order Buffer\xspace}
\newcommand{\ai}{AI\xspace}
\newcommand{\ailong}{Artificial Intelligence\xspace}

\newcommand{\plim}{PLIM\xspace}
\newcommand{\plimlong}{Programmable Logic in the Middle\xspace}
\newcommand{\api}{API\xspace}
\newcommand{\tme}{TME\xspace}
\newcommand{\tmelong}{Tensor Memory Engine\xspace}
\newcommand{\dtme}{D-TME\xspace}
\newcommand{\ctme}{C-TME\xspace}
\newcommand{\memory}{DRAM\xspace}
\newcommand{\pslong}{Processing System\xspace}
\newcommand{\pllong}{Programmable Logic\xspace}
\newcommand{\ps}{PS\xspace}
\newcommand{\pl}{PL\xspace}
\newcommand{\pspl}{PS-PL\xspace}
\newcommand{\pe}{PE\xspace}
\newcommand{\pelong}{Processing Element\xspace}
\newcommand{\soc}{SoC\xspace}
\newcommand{\soclong}{System-on-Chip\xspace}
\newcommand{\io}{IO\xspace}

\newcommand{\versal}{Versal\xspace}
\newcommand{\edge}{Edge+\xspace}

\newcommand{\axi}{AXI\xspace}
\newcommand{\ace}{ACE\xspace}
\newcommand{\tilelink}{TL\xspace}
\newcommand{\tilelinklong}{TileLink\xspace}
\newcommand{\wrap}{\texttt{WRAP}\xspace}
\newcommand{\id}{\texttt{ID}\xspace}
\newcommand{\axcache}{\texttt{aXcache}\xspace}

\newcommand{\lru}{LRU\xspace}
\newcommand{\lrulong}{\emph{least recently used}\xspace}
\newcommand{\mlp}{MLP\xspace}
\newcommand{\mlplong}{\emph{memory-level parallelism}\xspace}

\newcommand{\iid}{IID\xspace}
\newcommand{\iidlong}{independent and identically distributed\xspace}

\newcommand{\rdg}{RDG\xspace}
\newcommand{\rdglong}{Request Descriptors Generator\xspace}

\newcommand{\transnum}{N\xspace}
\newcommand{\latency}{latency\xspace}
\newcommand{\period}{period\xspace}
\newcommand{\freq}{frequency\xspace}
\newcommand{\config}{C\xspace}
\newcommand{\transaction}{T\xspace}
\newcommand{\data}{D\xspace}

\newcommand{\tmedimensions}{\ensuremath{N_{max}}\xspace}
\newcommand{\monitormlp}{\ensuremath{M_{max}}\xspace}
\newcommand{\pipelinenum}{P\xspace}
\newcommand{\targetmemmlp}{\ensuremath{L_{max}}\xspace}
\newcommand{\targetmemresp}{W\xspace}
\newcommand{\tmemaxdescriptors}{\ensuremath{D\xspace}}

\newcommand{\circled}[1]{\Circled[inner color=red]{#1}}

\newcommand{\para}[1]{\smallskip\noindent\textbf{#1}}
\newcommand{\objective}[1]{\para{Objective.}{#1}}

\newcommand{\paranoskip}[1]{\noindent\textbf{#1}}

\newcommand{\subpara}[1]{\textit{\underline{\smash{#1}}}}

\newtheorem{property}{Property}

\newcommand{\cube}[6]{
    \draw[#1, fill={#2}, opacity=0.2] ( #3+0, #4+0, #5-1) -- ++( 1, 0, 0) -- ++( 0, 1, 0) -- ++(-1, 0, 0) -- cycle; \draw[#1, fill={#2}, opacity=0.2] ( #3+0, #4+0, #5+0) -- ++( 0, 0,-1) -- ++( 0, 1, 0) -- ++( 0, 0, 1) -- cycle; \draw[#1, fill={#2}, opacity=0.2] ( #3+0, #4+0, #5+0) -- ++( 1, 0, 0) -- ++( 0, 0,-1) -- ++(-1, 0, 0) -- cycle; \draw[#1, fill={#2}, opacity=0.2] ( #3+0, #4+1, #5+0) -- ++( 1, 0, 0) -- ++( 0, 0,-1) -- ++(-1, 0, 0) -- cycle; \draw[#1, fill={#2}, opacity=0.2] ( #3+1, #4+0, #5+0) -- ++( 0, 0,-1) -- ++( 0, 1, 0) -- ++( 0, 0, 1) -- cycle; \draw[#1, fill={#2}, opacity=0.2] ( #3+0, #4+0, #5+0) -- ++( 1, 0, 0) -- ++( 0, 1, 0) -- ++(-1, 0, 0) -- cycle; \draw (#3+0.5, #4+0.5) node [anchor=center, align=center, font=\huge] {#6}; }

\definecolor{myorange}{RGB}{155, 155, 155} 
\maketitle

\begin{abstract}
    The shift to data-intensive processing from the cloud to the edge has introduced new challenges and expectations for the next generation of intelligent computing systems. As the compute-memory gap continues to grow, meeting these expectations implies that substantially more emphasis must be placed on careful memory optimization. Indeed, it is crucial for application-level access patterns to exhibit high spatiotemporal locality in caches using ideal memory layouts.
    However, only a minority of data-intensive applications are naturally characterized by ideal locality. Conversely, most applications exhibit either (i) poor locality when naively implemented, thus requiring costly redesigns and tuning, or (ii) an inflated memory footprint to offer proper locality. To address the aforementioned challenges, we propose to decouple computation from data layout (re)organization. We do so by introducing a Tensor Memory Engine (TME) capable of exporting dynamic alternative data layouts to CPU applications.
    
    Our approach transparently inserts TME on the CPUs’ data path when beneficial. TME performs on-the-fly data reorganization without data duplication to maximize cache locality by (i) accessing the memory on behalf of the CPUs and (ii) dynamically composing and serving re-organized cachelines to the upstream cache hierarchy. Unlike in- and near-memory computing approaches, it sets itself apart by clearly decoupling computation from memory (re)organization: computation is still performed on CPUs while data layout selection is delegated to the Tensor Memory Engine. TME follows a hardware/software co-design approach that can be implemented on commercially available SoC/FPGA platforms. We propose and evaluate a full-stack TME implementation on real hardware.
\end{abstract}

\section{Introduction}
    \label{sec:introduction}
    
    \para{Data Locality Remains a Fundamental Challenge.} The shift of data-intensive processing from the cloud to the edge has introduced new challenges and raised expectations for the next generation of intelligent computing systems. This trend has prompted many innovations across the hardware–software architecture continuum. In particular, significant effort has been devoted to devising sophisticated multi-level caching architectures to bridge the widening performance gap~\cite{wulf1995hitting,Patterson1997} between the \pelong{s} (\pe{s} such as CPUs, GPUs, and TPUs) and DRAM-based main memory.

    The effectiveness of caches, however, is dramatically affected by data access patterns and their degree of spatiotemporal locality.
    Overall, only a minority of data-intensive applications exhibit ideal locality.
    Instead, most applications that operate over high-dimensional data objects exhibit poor locality when na\"{i}vely implemented and must undergo costly redesigns and tuning to benefit from caching effects~\cite{tensors_caching}.

    \para{Data Layout of High-Dimensional Objects.}
    Many data-intensive applications naturally operate on high-dimensional objects, such as tables with many attributes, feature vectors, images, time series, or intermediate results in analytic pipelines~\cite{tensor_applications1}. Representing these structures as \textit{tensors} (\ie multi-dimensional arrays) provides a uniform abstraction for capturing both their logical organization (dimensions, indices) and their physical realization (layouts, strides, and tilings in memory)~\cite{tensor_applications2}. Achieving better cache locality when working with tensors often requires a tensor layout transformation: changing the order of dimensions, reshaping, blocking, or fusing/splitting axes corresponds directly to standard tensor operations such as transpose, reshape, and reindexing.
    The core challenge is that performing these transformations requires moving data through the memory hierarchy, often polluting it with unnecessary parts of the tensor.
    Moreover, these in-memory re-organizations of large tensors may not be possible on memory-limited devices (\eg edge systems).

    \para{Relationship with Existing In/Near Memory Processing.} Poor locality in several data-intensive applications has motivated research and investment in near- and in-memory computing~\cite{PIM21, near_mem_impl21}.
    Such approaches suggest moving part of the computation to where the data rest (\ie memory) to reduce data movement and alleviate the growing cost of traversing the memory hierarchy.
    However, such approaches often require specialized compute architectures and memory technologies, such as vendor-specific DRAM modifications---for example, Ambit provides in-DRAM bitwise operations with an estimated 1\% DRAM chip area overhead~\cite{Seshadri2017Ambit}. While these substantial architectural redesigns might be justified for specific popular applications, their highly customized designs restrict portability and adoption in commodity systems. As such, in this work, we study the possibility of mitigating the memory bottleneck problem while relying on traditional processor, cache, and main memory technologies. We do so by introducing additional hardware to offload memory re-organization and export \emph{ideal} locality without costly memory re-layouts and cache-aware algorithmic optimizations.

    \para{System-on-Chip to the Rescue.}
    In this paper, we posit that the democratization of \soc/\fpga offers a new avenue for near-/in-memory computing.
    We build on the idea advocated in~\cite{Roozkhosh2020, Roozkhosh2022CAESAR} of changing the role that the \fpga plays in the system: from hosting accelerators and peripherals to becoming part of the \pe{'s} data path.
    More precisely, we propose the \emph{\tmelong} (\tme); a cache-coherent \fpga-located module that electively handles memory requests from the \pe{s}' data path and, when doing so, responds with a re-organized representation of the memory that best fits the software's data locality.
    \tme achieves this by accessing the system's memory on behalf of the \soc{'s} \pe{s}; only expertly accessing the few memory locations needed to compose the requested re-organized cache line.
    This module enables the end-user to define \emph{any} re-organized view of dense data structures such as tensors without increasing the software memory footprint or working set size (WSS)---\ie without requiring the duplication of dense data structures in memory.
    
    \para{Software-Hardware Co-design for Ideal Locality.} The proposed \tme represents an answer to the question: \emph{Can we leverage application-level knowledge of data access patterns in the underlying memory hierarchy to export effortless and ideal data locality?}
    In other words, we focus on a key limitation of modern computing systems, \ie that the representation of data in micro-architectural caches mirrors the organization of data in main memory.
    Our work proposes an easy-to-program, easy-to-implement strategy for \emph{on-the-fly data transformation}, which is defined as the ability to re-organize, filter, compress, and reshape the content of data blocks as they are transferred from \ci their representation in main memory and \cii to \pe-facing caches.

    \para{Contributions.} Our work offers the following contributions.
    \begin{itemize}
        \item We describe and outline the architecture of \tme; a hardware/software co-design approach that aims to provide applications workloads with \ci ideal spatiotemporal cache locality and \cii limited increase of the WSS.
        \item We provide an open-source\footnote{Project URL omitted for review.} prototype of \tme that can be implemented on widely available commercial \soc/\fpga platforms; in our case the \amd \ultrascale Kria KR260 development board.
        \item We showcase, evaluate, and discuss the benefits of \tme using our prototype on various vision- and \ai-relevant operations on tensors. Our results indicate that reduced execution times and smaller memory footprints can be simultaneously achieved.
    \end{itemize}

 \section{Background}
    \label{sec:background}

In this section, we provide the necessary background on the hardware-software co-design concepts we build upon. 

\subsection{Tensors} 
    A tensor is a mathematical object that generalizes the concept of scalars, vectors, and matrices to higher dimensions. The number of dimensions of a tensor is referred to as the \emph{order}. Thus, an order-1 tensor is an array, while an order-2 tensor is a matrix. In vision as well as AI applications, tensors are used as the fundamental data structure to represent complex data objects like color images (3-order tensors) or video (4-order tensors). 
    A full tensor specification involves its \emph{shape}, which describes the specific dimensions of the tensor---\eg, a $1024\times1024\times3$ RGB image with a resolution of 1024 pixels by 1024 pixels. Additionally, tensors are homogeneous objects, and the \emph{data type} refers to the kind of value each element of the tensor holds---\eg, a floating point or integer. While modern AI frameworks such as TensorFlow~\cite{tensorflow2015-whitepaper} and PyTorch~\cite{pytorch} have popularized the association of tensors with neural networks, tensor processing is also at the core of traditional vision and signal processing pipelines~\cite{tensor_applications1}.

    \subsection{Processing In/Near Memory}
        Modern computing systems adhere to a compute-centric paradigm, in which data is transported to processing units for execution. This design choice is increasingly misaligned with contemporary workloads and technology trends: (1) applications are becoming more data-intensive and bandwidth-bound, and (2) data movement now dominates both energy consumption and performance overhead in modern platforms \cite{mutlu2025PIM}. This is exacerbated by the fact that DRAM scaling has failed to keep pace with the growth of computational throughput
        
        To address these trends, there has been a proliferation of research into Processing in Memory (PIM) and Processing Near Memory (PNM) architectures that locate compute capabilities closer to data. By executing certain operations either within the memory arrays themselves (PIM) ~\cite{pim_samsung_industrial_isca2021, pim_simdram_asplos2021} or in close proximity (PNM) ~\cite{pnm_past_present_future}, these approaches aim to alleviate bandwidth pressure on traditional memory hierarchies and improve performance for data-intensive workloads.

    \subsection{\pspl Platforms.}
        Since the early 2010s, the ever more heterogeneous computing landscape has seen the advent and nascent adoption of the so-called \pspl platform.
        This type of platform tightly combines a typical \soc architecture (referred to as \pslong or \ps) with a dynamically reprogrammable logic fabric (referred to as \pllong or \pl).
        In the vast majority of cases, the latter is implemented as a \fpgalong (\fpga).
        Many \pspl architectures have been designed by industry~\cite{zcu-102, intel_stratix10, microsemi_polarfire} and research-led~\cite{10121294, enzian2020cidr} projects.
        Each comes with its own specificity; however, the same key concepts prevail.
        In this paper, we assume a \soc architecture similar to Xilinx UltraScale+~\cite{zcu-102}.

        The \ps is composed of a multi-core computing cluster whose cores are connected via a \llclong.
        In addition to ensuring inter-core coherence, the \llclong (\llc) interfaces the core cluster with the memory subsystem.
        The \llc is connected to a DRAM controller via a coherent interconnect that routes requests across the \soc and, if enabled, ensures coherence between the core cluster and the \fpga.
        Note that the memory subsystem is located on the \ps side.
        The interfacing means between the \ps and \pl can be grouped into two categories:

\begin{enumerate}[leftmargin=15pt,itemsep=0pt,partopsep=0pt]
            \item [(i)] \textbf{Unidirectional high-performance interfaces.}
        \pspl platforms offer a set of unidirectional ports that allow communication (1) from \ps to \pl and (2) from \pl to \ps.
        Each \ps-to-\pl port is mapped to a unique \soc-wide physical address range, allowing any \pe-originated transaction to be explicitly routed to the \pl side.
        Similarly, \pl-to-\ps ports allow the \pl to access any memory target (\eg on-chip memories, DRAM controller) located in the \ps using their predefined addresses.
        Naturally, \pspl platform are agnostic to the bus protocol used but, the most popular and available \soc{s} implements the AXI-Full protocol~\cite{arm-axi-ace}.
        \item [(ii)] \textbf{Two-way coherent interfaces.}
        In some \soc models, the \pl is not confined to host peripheral modules and non-coherent accelerators.
        It can be added to the \soc coherence domain by enabling the coherence port.
        This enables the \pl to receive \emph{snoops} from the CCI, effectively exporting information about the memory-related activity occurring in the PS to the PL.
        Previous research~\cite{Roozkhosh2022CAESAR} exemplifies the level of detail that can be extracted and exploited this way.
        \end{enumerate}

    \subsection{Programmable Logic in the Middle}
        \label{subsec:}
        
        The Programmable Logic in the Middle (\plim) approach~\cite{Roozkhosh2020} describes a new way to employ the reprogrammable fabric of \pspl platforms.
        Instead of seeing the \pl side as solely a host for accelerators and peripherals controllers, \plim proposes to utilize the \pl as part of the \ps{'s} \pe{s}' data path.
        The key idea is that, by giving the \pl observability of the \ps{'s} memory activity, the \pl-located \plim module is able to analyze, manipulate, and monitor individual transactions.
        In~\cite{Roozkhosh2020}, \ps-originated memory transactions are routed to the \pl by instrumenting the software to explicitly target the address ranges of the unidirectional high-performance interfaces.
        Once in the \pl, the transactions are analyzed, manipulated, or monitored before being relayed to main memory via a simple address translation.
        Such approach was successfully used to \ci work around address-coloring side effects~\cite{Roozkhosh2020}, \cii highlight timing anomalies~\cite{9470445}, and \cii re-organizing database representation on-the-fly~\cite{Roozkhosh2022}.

        \para{Coherence Backstabbing.} \plim was later extended to leverage the existing two-way \pspl coherence interface to insert itself on the back of the coherence protocol, an approach called \emph{coherence backstabbing}~\cite{Roozkhosh2022CAESAR}.
        Doing so elevates the position of the \pl in the system and increases its privilege level and its level of observability.
        By being tethered to the \pspl coherence port, a \pl-located \emph{``backstabbing module''} is made part of the coherence domain.
        As such, its interaction with the \ps changes.
        Typically, following the coherence protocol, the backstabbing module is snooped by the \cci upon a \ps-located \llc refill.
        Then, the \cci expects the module to indicates whether it possesses a local copy of the requested cache line.
        The key idea of coherence backstabbing is to design a module that can \ci unilaterally decided to pretend possessing a local copy of the requested cache line and \cii answer with \emph{manufactured} data.
        The data can be either made up or originating from another memory source (\ie \`a-la \plim) as demonstrated in~\cite{Roozkhosh2022CAESAR}. \section{Memory Semantic Transformation}\label{sec:memory_semantic_transformation}

We now introduce the notion of \textit{memory semantic transformation}, which allows arbitrary access mapping for multi-dimensional objects stored in memory.

    \para{Intuition.}
    In traditional computing paradigms, programmers must first \emph{commit} to a data layout and then write algorithms to account for said layout. Indeed, the data address calculation logic is intertwined with the very logic that captures algorithmic semantics. Moreover, since the same data objects can undergo different processing steps, it is often the case that the selected data layout is not the optimal one for every step. To achieve the desired performance level, algorithms then undergo costly optimization rounds, often involving temporary data re-layout. 

    In this paper, we propose to fundamentally decouple core algorithmic semantics from data access semantics. At an intuitive level, consider an algorithm that requires access to a set of data items. If algorithmic and data semantics are decoupled, algorithm designers can focus on describing the operation required on the following data item(s), assuming that semantically subsequent items are positionally sequential.

    To this end, we propose the following. First, data is placed in memory in a given/arbitrary layout, which is perhaps the most natural way to represent the data at hand. Next, a hardware engine, namely \tme, is given access to the raw memory object and to a \emph{access pattern specification} that encodes the structure of data accesses following high-level algorithmic semantics. Finally, the algorithm on the processor (CPU or otherwise) accesses subsequent items required for the processing at hand as if they were contiguously placed in memory. The responsibility of making said items appear contiguous upon access via on-the-fly re-organization is left to the \tme.

\para{Formalization.} Let us consider the interface that needs to be provided by the \tme in order to perform algorithmic/data decoupling. We refer to the data space visible to the high-level algorithm as exported by the \tme as the \emph{reorganized data space}; we refer to the raw object accessed by the \tme as the \emph{non-reorganized data space}.

\tme must receive transaction requests targeting the reorganized data space. These transactions access $s$ bytes at an offset $o$ from an object-specific base address $a$.
We can therefore indicate the generic transaction as $\transaction_{a,o,s}$.
The object $a$ (and thus the transaction) is also associated with a pre-established \emph{access pattern specification} $\config$. 
Given these two elements, \tme must perform a scattered access within the non-reorganized data space to extract a data item $D_s$ of $s$ bytes.
If \tme is placed downstream from the processor-side caches, $s$ is equal to the cache line size. Extracting $D_s$ given the input transaction $\transaction_{a,o,s}$ is divided into three subproblems, namely (1)~\emph{address decomposition}, (2)~\emph{memory access}, and (3)~\emph{result aggregation}.

The \textbf{address decomposition} is the process of translating the transaction $\transaction_{a,o,s}$ in the reorganized data space into a series of smaller transactions into the non-reorganized data space following the specification $\config$: 
    \begin{equation}
        f_{decomp}~:~\transaction_{a,o,s}\times\config\mapsto\{\transaction_{b,o_0,s'},\ldots,\transaction_{b,o_n,s'}\}.
    \end{equation}
    $\transaction_{a,o,s}$ is decomposed into a series of $n+1$ smaller requests of size $s' < s$ each and such that $s = s'\cdot(n+1)$. All these requests target an object in the non-reorganized data space with base address $b$, while the various offsets $o_0,\ldots,o_{n}$ are derived from the specification \config as explained in the remainder of this section.

Next, \tme performs a series of parallel \textbf{memory accesses} following the decomposition step. This yields $n+1$ data fragments of size $s'$ each, thus:
    \begin{equation}
        f_{mem}~:~\{\transaction_{b,o0,s'},\ldots,\transaction_{b,o_n,s'}\}\mapsto\{\data_{s'_0},\ldots,\data_{s'_n}\}
    \end{equation}
Finally, \textbf{result aggregation} constructs a full cacheline of data $D_s$ from the individual fragments extracted in the previous steps, formally:
    \begin{equation}\label{eq:faggr}
        f_{aggr}~:~\{\data_{s'_0},\ldots,\data_{s'_n}\}\mapsto\data_{s}
    \end{equation}
It follows that the end-to-end behavior of \tme can be defined as the composition of the three steps:
    \begin{equation}
        f_{\tme}:\transaction_{a,o,s}\times\config\mapsto\data_{s} = (f_{decomp} \circ f_{mem} \circ f_{aggr})(\transaction_{a,o,s}, \config)
    \end{equation}

\para{Access Pattern Specifications.} Any object in reorganized data space is associated with an access pattern specification $\config$ that determines how subsequent reorganized data items map to data items in the non-reorganized data space. In this paper, we propose specifications for multi-dimensional stride patterns that can encode any repeating access pattern within a homogeneous multi-dimensional array (tensor).

An ($N+1$)-dimensional access pattern specification is an ordered set of $N+1$ tuples of the form:
    \begin{equation}
        \config:(\omega_N, \sigma_N, w_N),\ldots,(\omega_0, \sigma_0, w_0),
    \end{equation}
where each tuple describes a \emph{move} in the $i$-th dimension with $i\in\{0,\ldots,N\}$. Specifically, $\omega_i$ represents an initial offset to apply on the $i$-th dimension; $\sigma_i$  represents the size of an increment along the $i$-th dimension; and $w_i$ corresponds to the length of the $i$-th dimension.

By employing this specification, the \tme can translate any linear offset $o$ from the base address $a$ in the incoming transaction $\transaction_{a,o,s}$ into a set of per-dimension offsets $c_i$ as follows:
    \begin{equation}\label{eq:dim_counters}
        c_i = \omega_i + \bigg(o/\prod^{i-1}_{j=0}w_j\bigg)\%w_i,
    \end{equation}
where "/" and "\%" denote the integer division and modulo operation, respectively.
Finally, the first chunk of data to be retrieved from the non-reorganized data space will have offset $o_0$ from the base address $b$ that can be computed as:
    \begin{equation}\label{eq:nrd_off}
        o_0 = \sum^N_{j=0} c_i \sigma_i.
    \end{equation}
If the original transactions requested $s$ bytes while the size of an individual element in the non-reorganized tensor has size $s'$, such that $s/s' = n+1$, the remaining $n$ offsets $o_1,\ldots,o_n$ can be easily computed. This is done by iteratively incrementing the per-dimension offsets $c_i$'s while respecting the dimension lengths $w_i$'s and then re-computing Eq.~\ref{eq:nrd_off}.

    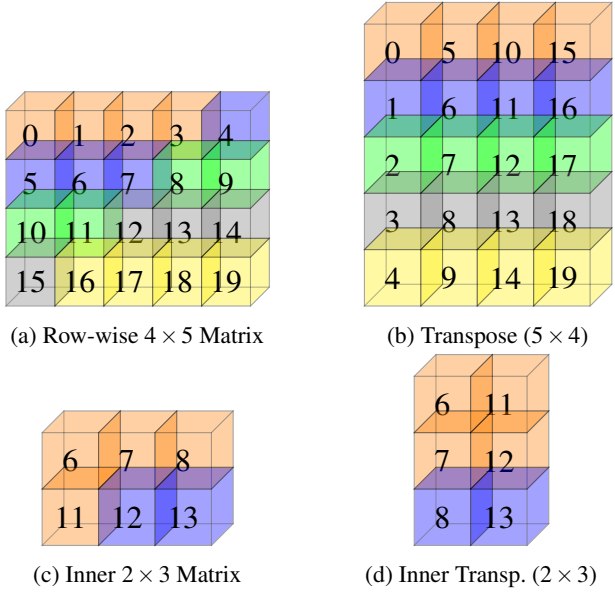
\begin{figure}
        \centering
        \begin{subfigure}[b]{0.45\columnwidth}
            \centering
            \begin{tikzpicture}[yscale=0.65,xscale=0.65]
    \cube{black!100}{orange}{0}{3}{0}{\large 0}
    \cube{black!100}{orange}{1}{3}{0}{\large 1}
    \cube{black!100}{orange}{2}{3}{0}{\large 2}
    \cube{black!100}{orange}{3}{3}{0}{\large 3}
    \cube{black!100}{blue  }{4}{3}{0}{\large 4}
\cube{black!100}{blue  }{0}{2}{0}{\large 5}
    \cube{black!100}{blue  }{1}{2}{0}{\large 6}
    \cube{black!100}{blue  }{2}{2}{0}{\large 7}
    \cube{black!100}{green }{3}{2}{0}{\large 8}
    \cube{black!100}{green }{4}{2}{0}{\large 9}
\cube{black!100}{green }{0}{1}{0}{\large 10}
    \cube{black!100}{green }{1}{1}{0}{\large 11}
    \cube{black!100}{gray  }{2}{1}{0}{\large 12}
    \cube{black!100}{gray  }{3}{1}{0}{\large 13}
    \cube{black!100}{gray  }{4}{1}{0}{\large 14}
\cube{black!100}{gray  }{0}{0}{0}{\large 15}
    \cube{black!100}{yellow}{1}{0}{0}{\large 16}
    \cube{black!100}{yellow}{2}{0}{0}{\large 17}
    \cube{black!100}{yellow}{3}{0}{0}{\large 18}
    \cube{black!100}{yellow}{4}{0}{0}{\large 19}
\end{tikzpicture}             \caption{Row-wise $4 \times 5$ Matrix}
            \label{fig:y equals x}
        \end{subfigure}
        \hfill
        \begin{subfigure}[b]{0.45\columnwidth}
            \centering
            \begin{tikzpicture}[yscale=0.75,xscale=0.75]
    \cube{black!100}{orange}{0}{4}{0}{\large 0}
    \cube{black!100}{orange}{1}{4}{0}{\large 5}
    \cube{black!100}{orange}{2}{4}{0}{\large 10}
    \cube{black!100}{orange}{3}{4}{0}{\large 15}
    
    \cube{black!100}{blue  }{0}{3}{0}{\large 1}
    \cube{black!100}{blue  }{1}{3}{0}{\large 6}
    \cube{black!100}{blue  }{2}{3}{0}{\large 11}
    \cube{black!100}{blue  }{3}{3}{0}{\large 16}
    
    \cube{black!100}{green }{0}{2}{0}{\large 2}
    \cube{black!100}{green }{1}{2}{0}{\large 7}
    \cube{black!100}{green }{2}{2}{0}{\large 12}
    \cube{black!100}{green }{3}{2}{0}{\large 17}
    
    \cube{black!100}{gray  }{0}{1}{0}{\large 3}
    \cube{black!100}{gray  }{1}{1}{0}{\large 8}
    \cube{black!100}{gray  }{2}{1}{0}{\large 13}
    \cube{black!100}{gray  }{3}{1}{0}{\large 18}
    
    \cube{black!100}{yellow}{0}{0}{0}{\large 4}
    \cube{black!100}{yellow}{1}{0}{0}{\large 9}
    \cube{black!100}{yellow}{2}{0}{0}{\large 14}
    \cube{black!100}{yellow}{3}{0}{0}{\large 19}
\end{tikzpicture}             \caption{Transpose ($5 \times 4$)}
            \label{fig:y equals x}
        \end{subfigure}
        \hfill
        \begin{subfigure}[b]{0.45\columnwidth}
            \centering
            \begin{tikzpicture}[yscale=0.75,xscale=0.75]
    \cube{black!100}{orange}{0}{1}{0}{\large 6}
    \cube{black!100}{orange}{1}{1}{0}{\large 7}
    \cube{black!100}{orange}{2}{1}{0}{\large 8}
\cube{black!100}{orange}{0}{0}{0}{\large 11}
    \cube{black!100}{blue  }{1}{0}{0}{\large 12}
    \cube{black!100}{blue  }{2}{0}{0}{\large 13}
\end{tikzpicture}             \caption{Inner $2 \times 3$ Matrix}
            \label{fig:y equals x}
        \end{subfigure}
        \hfill
        \begin{subfigure}[b]{0.45\columnwidth}
            \centering
            \begin{tikzpicture}[yscale=0.75,xscale=0.75]
    \cube{black!100}{orange}{0}{2}{0}{\large 6}
    \cube{black!100}{orange}{1}{2}{0}{\large 11}
\cube{black!100}{orange}{0}{1}{0}{\large 7}
    \cube{black!100}{orange}{1}{1}{0}{\large 12}
\cube{black!100}{blue  }{0}{0}{0}{\large 8}
    \cube{black!100}{blue  }{1}{0}{0}{\large 13}
\end{tikzpicture}             \caption{Inner Transp. ($2 \times 3$)}
            \label{fig:y equals x}
        \end{subfigure}
        \hfill
\caption{Logic view of target matrix (a) stored row-wise in memory and occupying 5 cachelines (color-coded). Layouts visible through \tme when accessing (b) its transpose view, (c) its inner $3\times 2$ matrix, and (d) the transpose of the inner $3\times 2$ matrix.}
        \label{fig:matrix}
    \end{figure}
    
    \para{Concrete Examples.}
        To better understand \tme specifications and its operation, consider a (2-dimensional) $4 \times 5$ matrix stored in the row-wise format in main memory in consecutive addresses starting at base address $b$, as per \fig[(a)]{fig:matrix}.
        Assume that each cacheline (color coded) includes $s = 4$ data items of some base data item size $s'$. This could be 1 for \texttt{int8}, 4 for \texttt{int32} and so on.

        Consider now the trivial case in which the processor wants to access the matrix in the same way in which it is stored using \tme. In this case, the reorganized data space base address $a_1$ is registered along with the specification $\config_1 = (0, 1, 20)$. The 1-dimensional stride $\config_1$ accesses the matrix linearly. For instance, the first cacheline request $T_{a_1, 0, 4}$ will be decomposed into $T_{b, 0, s'}, T_{b, 1, s'}, T_{b, 2, s'}, T_{b, 3, s'}$.
        
        More interestingly, consider the case in which the processor wants to access the transpose of the matrix while keeping the data in the row-wise format. This can be accomplished by registering a new address $a_2$ with configuration $\config_2 = (0, 1, 4),(0, 5, 4)$. In this case, $T_{a_2, 0, 4}$ will be decomposed into \{$T_{b, 0, s'}, T_{b, 5, s'}, T_{b, 10, s'}, T_{b, 15, s'}$\}, next $T_{a_2, 4, 4}$ into \{$T_{b, 1, s'}, T_{b, 6, s'}, T_{b, 11, s'}, T_{b, 16, s'}$\}, and so on, as shown in \fig[(b)]{fig:matrix}.

        By increasing the dimensionality of the stride, more complex patterns can be obtained. For instance, one can access the $2 \times 3$ sub-matrix in the center of the original matrix (\fig[(c)]{fig:matrix}). This can be done with $\config_3 = (1,5,1),(1,1,1),(0,5,2),(0,1,3)$ which yields the following transaction decomposition for the first cacheline access: \{$T_{b, 6, s'}, T_{b, 7, s'}, T_{b, 8, s'}, T_{b, 11, s'}$\}. Similarly, the transposed inner matrix depicted in \fig[(d)]{fig:matrix} can be accessed with the configuration $\config_4 = (1,5,1),(1,1,1),(0,1,3),(0,5,2)$.

        In the following sections, we describe the design and implementation of \tme, describing in greater detail how access pattern specifications are encoded; how processor-side requests are intercepted; and how decomposition, memory access, and result aggregation are efficiently implemented.

  \section{System Overview}
    \label{sec:system_overview}

    \begin{figure}[t]
        \centering
        \includegraphics[width=1.0\columnwidth]{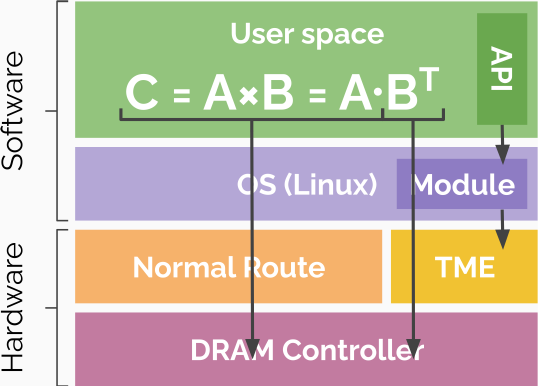}
        \caption{Overview of the system organization.}
        \label{fig:system_overview}
    \end{figure}

    In this section, we describe the software and hardware layers of the proposed system. The software layer is implemented as a combination of Linux kernel-level drivers and user-space libraries. The hardware layer is implemented in the PL and seamlessly interacts with the PS-side caches by leveraging the \plim paradigm---see Section~\ref{sec:background}.

    \begin{figure}[t]
        \centering
        \includegraphics[width=1.0\columnwidth]{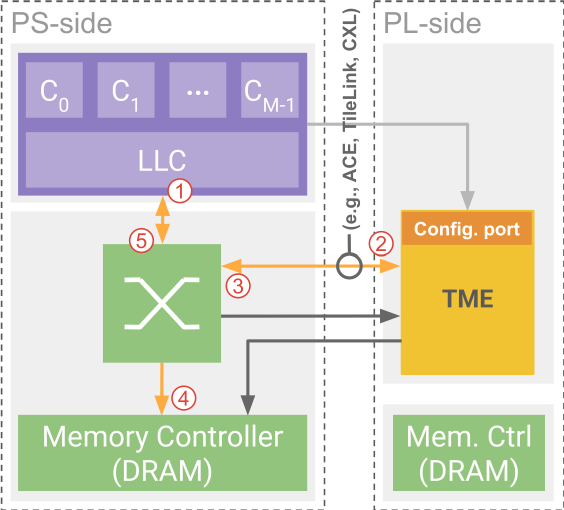}
        \caption{Abstract overview of \tme's integration in a \pspl platform as part of the coherent domain and accessed only upon \tme's decision. No assumptions on the bus protocol (\eg \ace, \axi) and \pl's location (\eg on- or off-chip) are made.}
        \label{fig:system_overview_tme}
    \end{figure}

    \para{Software layer.}
        We propose a set of \api{s} that allow end users to leverage \tme easily and without specialized knowledge.
        Said \api is intentionally made opaque to abstract away hardware-specific technical details.
        This is achieved by implementing a two-layer user/kernel software stack.

        The highest level of abstraction, in the user-space, provides end-users a way to describe dense data structures and target re-organization specification(s) in a form that closely follows that presented in Section~\ref{sec:memory_semantic_transformation}.
        
        The lowest level of abstraction corresponds to the \tme driver implemented as a Linux kernel module.
        The driver communicates directly with the \tme (\ie hardware layer).
At the driver level, the user configuration is translated into a \emph{``hardware specific''} and \emph{``hardware friendly''} representation.

        As illustrated in \fig{fig:system_overview}, the routing of memory transactions through the \tme module is elective.
        It means that accesses that do not require re-organization can reach memory using the usual data path, whereas, when explicitly deemed appropriate by the end-user, selected tensors can be presented to the \pe{s} as re-organized by accessing the \tme \api.
        The latter will ensure the re-routing of the \pe-originated traffic through the \tme module.
Let us consider the example in \fig{fig:system_overview} of a matrix multiplication $C = A \times B$ where the matrices $A$ and $B$ are stored as row-major.
        Only $B$ needs to be re-organized (here transposed) to allow the matrix multiplication to be performed as a dot product.
        Further details on the \api are provided in Section~\ref{sec:memory_semantic_transformation}.

    \para{Hardware layer.}
\tme is integrated within the system cache coherence via an interface linking the \pl with the \ps coherence fabric (\eg \ace, \tilelinklong).
        \fig{fig:system_overview} illustrates such integration.
        The \pe{'s} requests emitted in \circled{1} are sent to all coherent actors following the snoop-based coherence mechanism \circled{2}.
        \tme decides whether to handle the request and informs the coherent fabric in \circled{3} of its decision.
        If accepted, \tme sends the requests to the memory as non-cached through the coherent interface, effectively accessing the data using the usual data path \circled{4}.
        Once the re-organized cacheline is ready, it is forwarded to the \pe{s} via the coherent interface \circled{5}.
        Overall, \tme centralizes all memory interaction on a single interface.

        Note that \tme can also be integrated as a memory-mapped target.
        However, in such case, the selectiveness aspect is lost, and two ports are required (a subordinate and a main) as the module becomes a \emph{``passthrough''}.
     \section{\tme Architecture}
    \label{sec:architecture}

    \begin{table}[t]
        \centering
        \caption{\tme's main architectural configuration parameters: shorthand notation and description.}
        \label{tab:architecture_parameters}
        \begin{tabular}{p{0.15\columnwidth}p{0.75\columnwidth}}
            \toprule
            Notation           & Description \\
            \midrule
            \tmedimensions     & Amount of dimensions that \tme can handle (\ie supports the re-organization). \\
            \monitormlp        & Amount of simultaneous outstanding transaction that can be handled by the Monitor. \\
            \targetmemmlp      & Memory-level parallelism of the memory targeted by \tme to fetch the re-organized cache-lines pieces. \\
            \tmemaxdescriptors & Amount of re-organization patterns simultaneously supported by a \tme implementation. \\
            \bottomrule
        \end{tabular}
    \end{table}

    We now present in detail the internal architecture of the proposed Tensor Memory Engine.
    For clarity, a summary of the architectural parameters is available in Table~\ref{tab:architecture_parameters}.

    \begin{figure}[t]
        \centering
        \includegraphics[width=1.0\columnwidth]{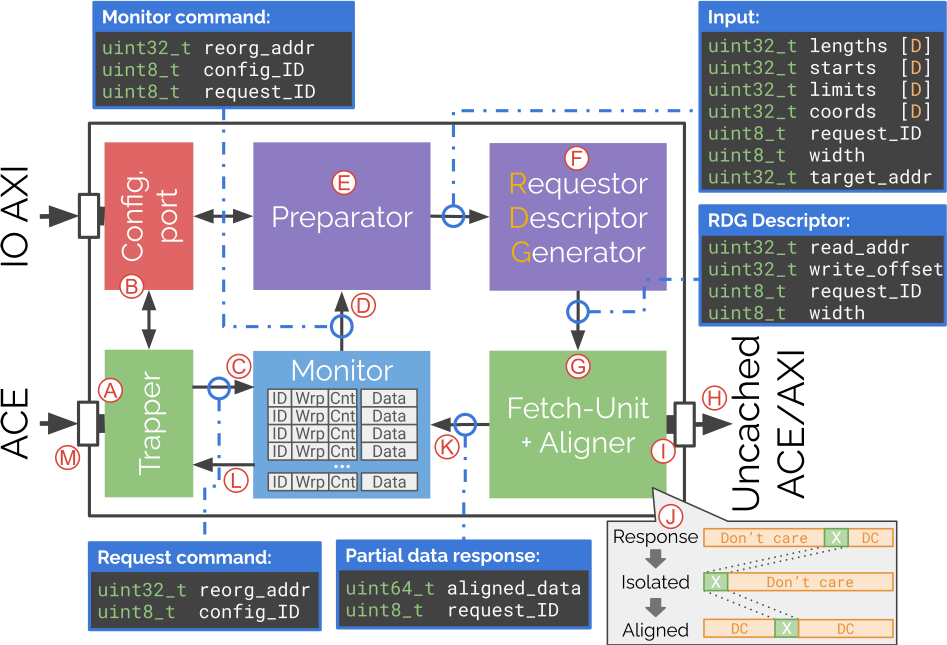}
        \caption{\tme's architecture overview. The \ace port is used for both capturing transactions of interest and (Snoop channels) and fetch data from memory (\axi channels). The performance monitoring unit is omitted for readability.}
        \label{fig:tme_arch_overview}
    \end{figure}

    \para{Overview.} The \tme architecture illustrated in \fig{fig:tme_arch_overview} is composed of eight modules and two bus ports.
    One of the ports is used by the \ps to indicate to \tme how to re-organize the data flowing through.
    In our prototype, this is achieved by writing with \io transactions to a set of \pl-located memory-mapped registers connected to the \ps via an \axi bus.
    The second port is in charge of intercepting transactions that fall within the memory ranges of interest.
    In our prototype, this port is coherent and included in the \ps coherency domain using the \ace protocol.

    \subsection{Components}
    
        \para{Configuration port.}
            The configuration port is a central component of \tme that stores all operands required to apply the manipulations described in Section~\ref{sec:memory_semantic_transformation}.
            It is composed of two arrays: a validity array and a descriptor array.
            Each array has a depth (noted \tmemaxdescriptors) equal to the amount of re-organization patterns simultaneously supported by \tme.
            The validity array is a set of booleans indicating whether the corresponding set of descriptor entries is ``in use''.
            The descriptor array stores the re-organization patterns as sets of \tmedimensions entries.
The arrays' content can be read and written via an \axi port tethered to one of the memory-mapped \pspl \axi interfaces.
Entries are \ci allocated and populated by the \api before any re-organized cache line requests are made, and \cii cleared by the \api when the re-organized accesses are completed.

        \para{Trapper.}
            The trapper is the first module on the data path.
            It receives and, when applicable, handles snoop commands (\ie transactions) coming from the \ps.
            The module determines whether to handle the transaction by comparing the transaction's address against a set of configured address ranges. Each range corresponds to a registered object in the re-organized data space.
            If the snoop address falls within one of the ranges, \tme informs the \ps that it will handle the request.
            Otherwise, it issues a \emph{``snoop miss''}.
            Upon handling a transaction, a few pieces of metadata---request address, \wrap, \id---are extracted and passed to the next module.
            The trapper module is also in charge of transforming the responses into transactions compliant with the \ace protocol.
        
        \para{Monitor.}
            The monitor module occupies a central role in the operation of \tme as it orchestrates the incoming and outgoing transactions.
            It receives the several pieces of a \emph{``re-organized cache-line''} in \emph{any order} but must respond to \pe-originated transactions with completed cache-lines in the exact order they arrived, effectively implementing result aggregation ($f_{aggr}$, see Eq.~\ref{eq:faggr}) as per Section~\ref{sec:memory_semantic_transformation}.
            The module operates as a \roblong (\rob) to respect the system's cache coherence interconnect. Indeed, the latter expects to receive in-order responses while \tme allows the out-of-order retrieval of partial data to achieve higher overall throughput.
            The \rob also stores the incoming transactions' information and is flanked with a data array that holds the partially constructed (\ie unfinished) re-organized cache-line(s).
            The monitor can simultaneously handle up to \monitormlp outstanding transactions.

        \para{Preparator.}
            This module is in charge of decomposing the re-organized data space offset $o$ of an incoming transaction $\transaction_{a,o,s}$ buffer into per-dimension offsets $c_0,\ldots,c_N$ required to fetch the various cacheline fragments within the non-reorganized data space.
            This operation corresponds to a hardware-optimized calculation of Eq.~\ref{eq:dim_counters}.
            Specifically, it is implemented as a pipeline to allow for the parallelization of the \tmedimensions successive division and modulo operations.
            For each operation, the preparator looks up and fetches its operands from the configuration port using the configuration \id retrieved by the trapper and passed along by the monitor.
                
        \para{Request Descriptors Generator.}
            As its name suggests, based on the set of dimension offsets $c_0,\ldots,c_N$ computed by the preparator, the \rdglong (or \rdg) generates a sequence of transaction descriptors\footnote{Note that the module uses descriptors to remain protocol agnostic.} that encode which memory location will be read. The first memory location corresponds to the base address of the non-reorganized object $b$ plus the offset of the first data item $o_0$, computed as per Eq.~\ref{eq:nrd_off}.      
The \rdg then generates one new sub-transaction descriptor every clock cycle, for $n$ clock cycles---once again, $n = (s/s' - 1)$. Thus, at each clock cycle, a new sub-transaction of the form $T_{b, o_i, s'}$ is formed and queued downstream at the Fetch Unit.
                
        \para{Fetch Unit.}
            The Fetch Unit module is in charge of \ci forming requests that can be transformed into well-formed bus transactions (\ie here \axi) based on the transaction descriptors produced by the \rdg. This step corresponds to requesting the data fragment $D_{s'_i}$ from the generic descriptor $T_{b, o_i, s'}$; \cii storing the request descriptors until the corresponding responses are delivered and \ciii, upon reception of these responses, retrieving the data of interest and aligning with its future location in the re-ordered cache-line, thus reconstructing the full $D_s$ for the original request $T_{a,o,s}$ originated from the processor. 

            As the target memory may provide different response times for each request\footnote{\eg if interacting with a caching hierarchy.}, the module must be capable to handle re-ordered individual fragments' responses.
            To this end, the module implements a \emph{transaction \id allocation table}.
            Typically, the module allocates a dedicated transaction \id to each incoming request descriptor and stores \rdg-generated metadata for later.
            Upon the reception of a transaction's response, the metadata is retrieved from the table using the \id of the response.
            The metadata is used to isolate the data of interest from the response and align it to its future location (\ie offset) within the re-organized cache line.
            The depth of the table is set to the same value as the targets' \mlp (noted \targetmemmlp).

        \subsection{Example: In-\tme Request Life Cycle}
            This subsection describes the \tme{'s} internal operations that occur between the reception of a snoop command and the transmission of the re-organized cache line to the \cci.
            The life cycle of a request within \tme is as follows.
            
            Upon a \llc refill (\circled{2} in \fig{fig:system_overview_tme} and Section~\ref{sec:system_overview}), the \cci snoops the \pspl coherence port, initiating interaction with \tme{'s} Trapper module \circled{A}.
            The latter determines whether to handle the request by looking up the configuration port's validity array \circled{B}.
            Providing a negative outcome, the \tme indicates the \cci that it does not want to handle the request (\ie it does not possess a local copy).
            Otherwise, \tme answers positively to the \cci and forwards the snoop request's address (\ie \texttt{reorg\_addr} in \fig{fig:tme_arch_overview}) to the Monitor \circled{C}.
            The Monitor \ci allocates an entry in the \rob for the new request, \cii stores collected metadata (\eg \wrap) in the \rob, and \ciii passes on the request to the Preparator \circled{D}.
            Hitherto completion, the entry's position within the \rob is internally used as an \id to identify the request (\ie \texttt{request\_ID} in \fig{fig:tme_arch_overview}).
            Together, the preparator and the \rdg translate the \texttt{reorg\_addr} into a sequence of \transaction descriptors (\ie \rdg descriptor in \fig{fig:tme_arch_overview}) as explained earlier \circled{E} \circled{F}.
            Each transaction descriptor is then \ci inserted in the Fetch-unit's allocation table \circled{G}, \cii transformed into an \axi transaction, and \ciii placed on the \ace port \circled{H} (\circled{4} in \fig{fig:system_overview_tme} and Section~\ref{sec:system_overview}).
            Upon reception of a response \circled{i}, the \rdg descriptor is retrieved from the table and used to isolate the data of interest from the response payload \circled{j}.
            Then, the result is forwarded to the Monitor, where the data is inserted in the partially constructed cache line and the associated counter incremented \circled{K}.
            Whenever the counter indicates a completed re-organized cache line, and that cache line is the oldest in the \rob, the cache line is forwarded to the Trapper \circled{L} and placed on the \ace bus \circled{M} (\circled{5} in Fig.~\ref{fig:system_overview_tme} and Sec.~\ref{sec:system_overview}).

    \section{Evaluation}
    \label{sec:evaluation}
    In this section, we evaluate the performance of TME utilizing a set of benchmarks representing common vision/AI and data manipulation workloads.

\subsection{Experimental Methodology} \label{sec:benchmarks}

\paranoskip{System Setup.} For hardware, we use the \amd \ultrascale Kria KR260 platform equipped with \ci a quad-core ARM Cortex-A53 CPU clocked at 1.3~GHz, \cii a tightly integrated programmable FPGA fabric, \ciii 1~MB of L2 cache, and \civ 4~GB of 2400~MT/s DDR4.
\tme is implemented on the \pl and operates at 300~MHz.
As discussed in Section~\ref{sec:architecture}, it interacts with the \ps via one \axi interface for configuration and one \ace interface for re-organized data accesses.

For the operating system, we use Ubuntu 22.04, provided by \amd~\cite{AMDInstallUbuntu} (Linux kernel v6.2), on top of which we run all our experiments. 
To provide easy access to performance counters, we employ the RT-Bench~\cite{rt-bench} framework when running the benchmarks.
Unless mentioned otherwise, all benchmarks are compiled with GCC v11.4.0, \texttt{-O2} optimizations, and auto-vectorization flags\footnote{\texttt{-funroll-loops -ftree-vectorize}} when applicable.

\para{Benchmarks.}
For benchmarks, we use a set of workloads that involve various \textit{tensor} manipulations, as described below.

\subpara{Im2col.} Convolutions underpin modern CNN workloads, motivating substantial effort to optimize both computation and data movement~\cite{demystifying-im2col, duplo, gpu-perf-mon}. A widely used implementation strategy is \textit{Im2col}, which reshapes the input feature maps so that convolution can be performed as a GEMM. While this approach leverages highly optimized linear-algebra kernels, it incurs significant data movement: for modest kernel sizes $(5\times5)$, the expanded matrix can exceed the input size by $\sim25\times$. TME eliminates this overhead by forgoing explicit materialization of the Im2col matrix. Instead, it generates the required layout on-the-fly and delivers data to the processing elements in precisely the order expected by the GEMM kernel. To illustrate the benefit, we evaluate a 1024×1024 grayscale input with a 2×2 convolution filter, comparing a conventional CPU Im2col implementation against a TME variant that constructs the convolution matrix on demand.
    
\subpara{Conv2D.} This benchmark is a naive implementation of 2D convolution with a single filter, where the filter is applied using a simple nested-loop strategy. On TME, this benchmark applies the convolution to a TME-assisted layout that ensures each subsequent datum is presented in sequential order, maintaining optimal data locality and effectively flattening the base image to match the convolution pattern. To enable comparison with Im2col, we again use a $1024\times1024$ grayscale input with a $2 \times 2$ convolution filter.

\subpara{Permutation.} A permutation operation reorders the axes---and thus the logical strides---of a tensor. Because strided accesses over a tensor result in irregular memory access, it is common to physically reshape the tensor to improve locality, albeit at the cost of substantial data movement, before processing it. 
In this benchmark, we permute a tensor of shape \((N, H, W, C)\), representing a batch of \(N\) images of size \(H \times W\) with \(C\) channels each. The tensor is re-shaped as a \((N, C, H, W)\) tensor, effectively removing color channel interleaving. We then apply a \(2 \times 2\) convolution filter as in Conv2D. 
We set \(N = 8\), \(C = 3\), and \(H = W = 512\). 
On TME, the permutation operation is carried out transparently on-the-fly. 
On the baseline CPU, however, the reordered tensor must be physically materialized in memory before being passed to the convolution kernel.
        
\subpara{Unfolding.} A tensor unfolding operation converts a higher-order tensor into a matrix by rearranging its elements such that one axis becomes the rows of the result and all remaining axes collapse into columns. 
For example, consider a tensor $\chi \in \mathbb{R}^{2 \times 3 \times 4}$. 
A mode-1 unfolding of $\chi$ yields the shape $(2, 12)$, a mode-2 unfolding yields the shape $(3, 8)$, while a mode-3 unfolding yields the shape $(4, 6)$. More generally, for an $N$-order tensor, $\chi \in \mathbb{R}^{I_1 \times I_2 \times \ldots \times I_N}$, the mode-$k$ unfolding yields $\chi_{(k)} \in \mathbb{R}^{(I_k) \times (I_1 \times I_2 \times \ldots \times I_{k-1}\times I_{k+1} ... \times I_N)}$. 
For this experiment, we consider a tensor $\chi_{1} \in \mathbb{R}^{8 \times 64 \times 64 \times 128}$ and a 2-order tensor (matrix) $\chi_{2} \in \mathbb{R}^{64\times 65536}$. We first unfold $\chi_{1}$ in mode-3 such that its dimensions align with those of $\chi_{2}$, and then compute the Hadamard product~\cite{tensor_applications2} between the resulting matrices. The unfolding of $\chi_{1}$ is carried out either explicitly on the CPU or implicitly on-the-fly by TME. 

\subpara{Batch2Space.} This operation is applied to a tensor whose batch dimension encodes subdivided spatial data. It rearranges this data by moving elements out of the batch dimension and into the spatial dimensions, producing an output tensor with enlarged spatial dimensions and a reduced batch size. This operation is common in real-world ML and signal processing workflows~\cite{batch2space, tensor_applications1}. 
For this experiment, we instantiate an image batch with dimensions $(N, H, W, C)$, where $N=8$, $H=W=64$, and $C=3$. We perform the \emph{Batch2Space} operation to collapse them into a single $128\times256$ image. This is done na\"lively by moving data with the CPU, and then using on-the-fly re-organization via TME. A $2\times2$ convolution filter is then applied to the resulting image.

\subpara{MatMul.}
In this benchmark, we multiply two $2048\times2048$ square matrices. We pre-transpose the second matrix either via the CPU or through a TME-assisted re-layout. For multiplication, we utilize a recursive divide-and-conquer strategy~\cite{clrs}. 

\subpara{Slicing.} 
Tensor \emph{Slicing} is a multi-dimensional generalization of array slicing. For this experiment we take a base tensor $\chi \in \mathbb{R}^{R_1 \times R_2 \times R_3 \times R_4}$. We perform a strided slice of the base tensor to get $\chi' \in \mathbb{R}^{(R_1/2) \times (R_2/4) \times (R_3/2) \times (R_4/64)}$. This is achieved by accessing along each dimension, starting with a null offset, with strides $(S_1, S_2, S_3, S_4) = (2, 4, 2, 64)$. For our experiments, we set $(R_1, R_2, R_3, R_4) = (64, 64,64, 512)$.
We then take a Hadamard product of $\chi'$ and a secondary tensor with the same shape. We perform this operation via TME and compare it with in-place access on the CPU. In this case, we found that the TME approach was much faster than explicitly materializing the sliced tensor.

\begin{figure*}[htp]
  \centering
  \begin{subfigure}[t]{0.48\textwidth}
    \centering
    \includegraphics[width=\linewidth]{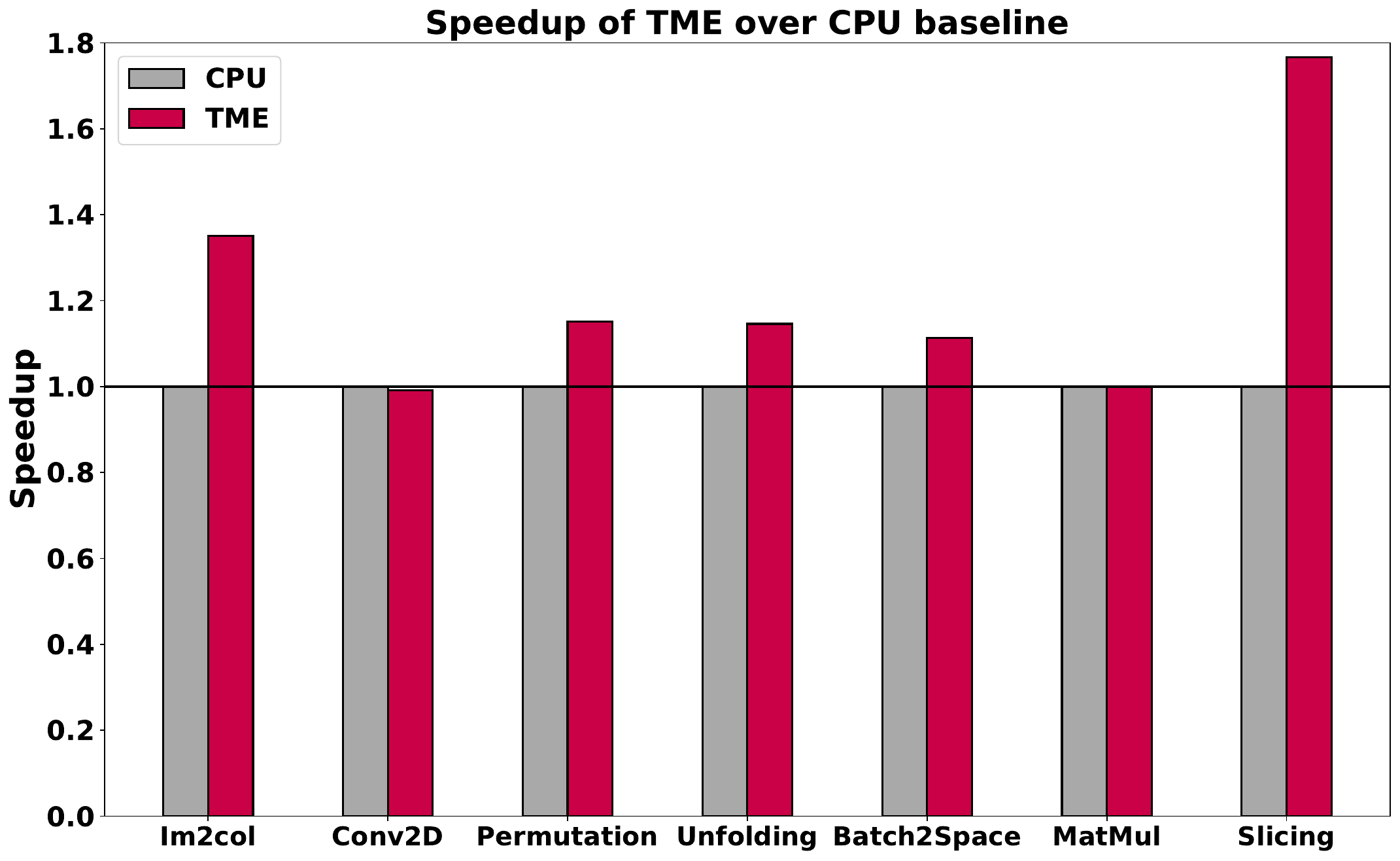}
    \caption{Speedup of TME benchmarks relative to their CPU counterpart}
    \label{fig:tensor_op}
  \end{subfigure}
  \hfill
  \begin{subfigure}[t]{0.48\textwidth}
    \centering
    \includegraphics[width=\linewidth]{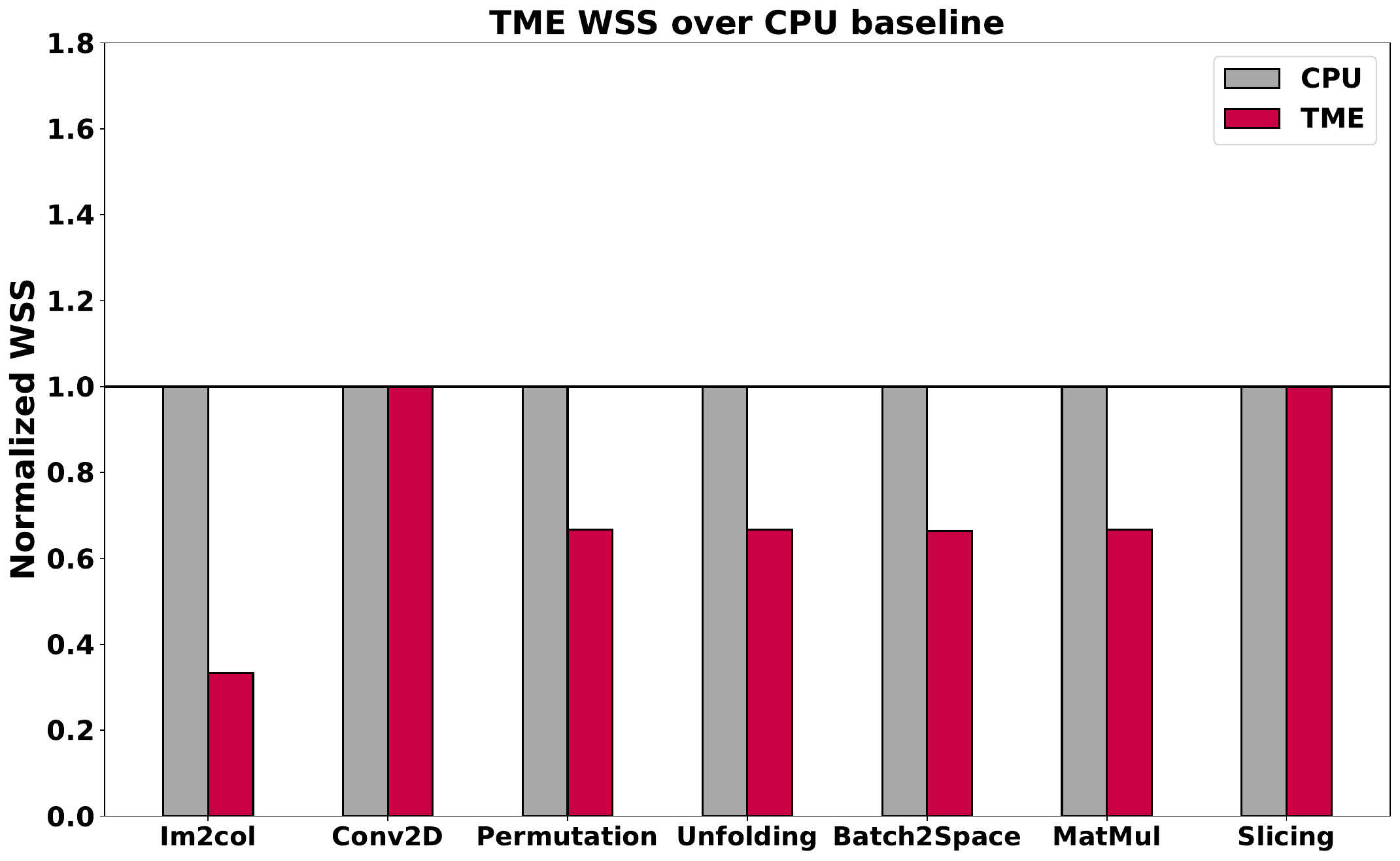}
    \caption{TME benchmark WSS normalized to their CPU counterpart}
    \label{fig:tensor_op_memory}
  \end{subfigure}
  \caption{\tme is able to offer \ci speedup by eliminating transformation time (\fig{fig:tensor_op}) and \cii ideal locality decrease transient working set size (WSS) by avoiding materialization of intermediate layouts (\fig{fig:tensor_op_memory}).}
  \label{fig:tensor_benchmark}
\end{figure*}

\subsection{Speedup and Working Set Size Reduction}

We first evaluate \tme{'s} effects on performance and memory footprint using the aforementioned tensor manipulation benchmarks.
\fig{fig:tensor_benchmark} shows the results.

\para{Speeding Up Tensor Layout Transforms by Avoiding Materialization.}
    \emph{Im2col}, \emph{Conv2D}, \emph{Batch2Space}, \emph{Unfold}, \emph{Permutation}, and the transpose in \emph{MatMul} are all transformations that require a reorganization of large tensors into new views, which in the baseline implementation results in substantial memory movement and a significant increase in transient WSS.
    \tme improves the performance of these workloads by eliminating the materialization cost of these transient tensor views and by exposing the desired layout directly to the CPU.
    This means that the explicit transformation operations disappear entirely from the critical path: no intermediate buffer is allocated, filled, or written back.
    As a result, \emph{Unfold}, \emph{Permutation}, and \emph{Batch2Space} see a $1.15\times$, $1.15\times$, and $1.11\times$ speedup relative to their CPU-only baseline, while \emph{Im2Col} shows the largest improvement at $1.35\times$---see \fig{fig:tensor_op}.
    In the case of \emph{MatMul}, the $O(n^3)$ matrix multiplication time dominates the time of the $O(n^2)$ transpose, and little-to-no benefit is seen in the execution time, but we still see a significant decrease in WSS.
    The benefit of \tme is particularly significant for \emph{Im2Col} because the output representation is much larger than the input tensor, and the materialization cost often dwarfs the compute time of the succeeding GEMM.
    By eliminating this inflated intermediate representation, \tme reduces pressure on the memory hierarchy, reduces WSS, and prevents bandwidth from being monopolized by data rearrangement, as shown in \fig{fig:tensor_op_memory}.

\para{TME Improves Slicing by Converting it to a Streaming Access Pattern.}
\emph{Slicing}
achieves a $1.77\times$ speedup relative to the CPU baseline. In this case, the WSS does not improve because, as mentioned in Section~\ref{sec:benchmarks}, instead of materializing an intermediate tensor, we compare to a faster in-place CPU benchmark. \emph{Slicing} is fundamentally different from the rest of the benchmarks in that it selects only a subset of the base tensor, rather than rearranging it. The operation's strided accesses exhibit worst-case spatial locality, leading to extremely poor cache line utilization. Only a small portion of the retrieved lines contains data that will be used, resulting in a severely inefficient use of memory bandwidth and cache space. Furthermore, because the strides are nested, the access pattern exhibits low prefetchability and fails to form a stable stream, leading to increased memory access latency. With TME, both of these issues are addressed simultaneously. TME can compose and present lines filled with useful data to the cache hierarchy, achieving optimal locality. Additionally, by linearizing the useful data, nested multi-dimensional strided accesses are eliminated, making the streams simple and prefetcher-friendly. This latter effect is particularly significant as it effectively hides the increased memory access latency incurred by TME~\cite{Strickler2025MicroArchitecturalER}. The synergy of improved cacheline utilization and high prefetchability enables TME-executed \emph{Slicing} to transition from a pathological access pattern to a streaming access pattern, yielding the largest speedup in the set.

\para{The Cost of Misaligned Tensor Layouts in Conv2D.}
Finally, \emph{Conv2D} sees a slightly worse performance relative to the baseline. In this experiment, while the TME-based approach applies a flattening procedure superficially similar to \emph{Im2col}, the resulting layout does not align well with the shape required by optimized GEMM routines. We effectively duplicate data items and, because of their format, cannot operate on them with SIMD instructions. 
As such, the unfolded layout forces the system to move a larger volume of data across the memory hierarchy per operation, increasing the bandwidth consumed along the TME–DRAM path. This negative result indicates that proper algorithm co-design principles must be adhered to when using TME. Next, we examine the bandwidth limitations along the TME-DRAM path in more detail.

\begin{figure}[t]
    \centering
    \includegraphics[width=1.0\columnwidth]{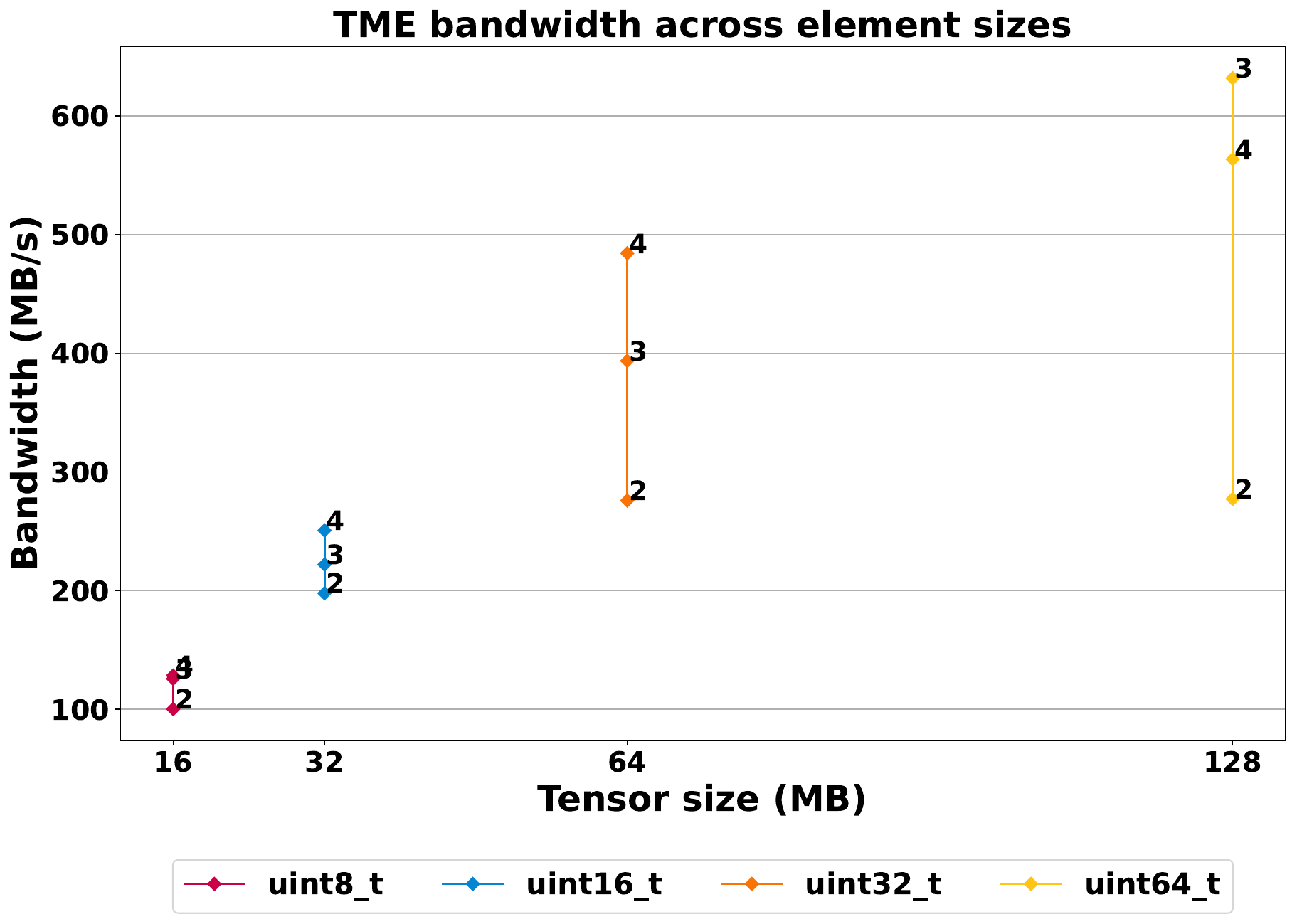}
    
    \caption{For smaller element sizes, more TME-DRAM transactions are necessary to compose a cache line, effectively limiting bandwidth along the TME-DRAM path.}
    \vspace{-0.85em}
    \label{fig:bandwidth_synthetic}
\end{figure}

\para{TME as a Request Multiplier: DRAM Bandwidth Implications.}
We now investigate the effective memory bandwidth between TME and DRAM. We utilize the \emph{Bandwidth}~\cite{isol-bench} benchmark, which sequentially accesses a large array, to point to a TME-provided layout with various element sizes. \fig{fig:bandwidth_synthetic} shows the results. 
Note that the in-graph annotations denote the dimensionality of each TME configuration.

TME effectively functions as a request multiplier: a single upstream cache line request is expanded into multiple fine-grained scatter-gather accesses that are composed into a contiguous cache line. Critically, DRAM devices are optimized for cache line–sized accesses. An AXI read request from TME for a sub-cache line payload (\eg 8 bytes) is not distinguished by the DRAM circuitry. Instead, it is handled internally as a full-burst access, with granularity effects only enforced when the memory controller returns data in accordance with the AXI specification. This means that although TME extracts less data, it still incurs the full latency cost. The number of transactions required by TME to compose a 64-bytes cache line is equal to $n =\frac{64 bytes}{element\_size}$. For small element sizes (\eg \texttt{uint8}/1~byte), 64 separate transactions to DRAM are required for TME to compose a single cache line. Each of these transactions must incur the normal DRAM latency, leading to a substantial reduction in achievable bandwidth on the TME-DRAM path.  As the element size increases, fewer total transactions are needed to compose a cache line, and TME-DRAM bandwidth increases.
This bandwidth limitation delineates the situations in which TME can be used effectively and must be accounted for in the algorithm co-design phase. Importantly, this bandwidth degradation is not inherent to on-the-fly data reorganization as a paradigm, but rather an artifact of interfacing with commodity DRAM controllers optimized for cache line granularity.

\para{Summary.}
TME enables the CPU to consume tensor data in the desired logical format without materializing an intermediate data layout. This eliminates the storage and latency overheads associated with explicit data transformation. Beyond eliminating in-memory materialization of reorganized objects, TME fundamentally reshapes memory behavior by converting irregular, multidimensional access patterns into linearized streams that align properly with cache hierarchies and simple hardware prefetchers typical of edge nodes. 

Our evaluation demonstrates that while many tensor transformations benefit from removing materialization overhead, the largest gains arise when TME improves memory-system efficiency by maximizing cacheline utilization and prefetch coverage. These results highlight TME’s potential as a general hardware–software co-design for accelerating data-intensive workloads whose performance is dominated by memory access behavior rather than computation.

        \section{Discussion}
    \label{sec:discussion}

One of our observations from our experiments is that TME effectively acts as a request multiplier for some workloads on commodity DRAM, where a single cache-line request from the CPU is decomposed into multiple sub-cache-line gathers, each of which still incurs full-burst cache-line size access to DRAM. For small element sizes, this behavior can reduce effective bandwidth on the TME–DRAM path, even though the data reorganization itself is conceptually efficient. We next discuss a possible avenue for mitigating this by co-designing TME’s access patterns with the underlying memory system.

\para{On-the-fly reorganization with new memory technologies}
Looking forward, these constraints could be substantially alleviated by adapting memory controllers and DRAM devices to support variable or finer-grained burst lengths. This direction aligns with recent trends in custom and near-memory technologies~\cite{mutlu2025PIM,hbm}, heterogeneous memory architectures~\cite{efficient-hetero-memory,page-placement-hetero,hetero-os}, and research on dynamic, fine-granularity DRAM access~\cite{dgms,sectored-dram}, as well as commercial designs with reduced burst lengths~\cite{RLDRAM3}. In future work, we plan to explore TME-like modules and on-the-fly data transformation under emerging memory technologies that better match TME’s accommodate patterns.

\para{RISC-V and ASIC Integration}
Another potential avenue to alleviate TME–DRAM bandwidth constraints is to integrate TME as an ASIC in a system-on-chip, allowing it to operate at a higher frequency and thereby improving its ability to extract more DRAM bandwidth. Toward this goal, we are actively exploring the integration of TME into an open-source RISC-V SoC. Such integration also opens the door to leveraging RISC-V’s extensible ISA for tighter architecture-level coupling with TME, which we plan to investigate in future work.

 \section{Related Work}
    \label{sec:related_work}

    \para{Data Locality and Tensor Layouts.}
    Systems such as TVM~\cite{chen2018tvm} focus on operator- and graph-level scheduling (tiling, fusion, vectorization, memory scoping) to produce hardware-efficient kernels across diverse backends.
    However, these systems ultimately realize layout transformations by explicitly materializing intermediate tensors or by carefully scheduled loops in software, thus still pushing all data reorganization through the conventional cache/memory hierarchy. In contrast, \tme moves layout transformation into the memory path itself, exporting locality-optimized cache lines while leaving the logical tensor program and CPU execution model unchanged.
    
    \para{Processing-In-Memory.}
    Processing-in-memory (PIM) and near-data computing reduce data movement by colocating simple computation with memory arrays or stacked memory~\cite{MutulRealWorldPIMTutorial}. DRAM-centric proposals such as Ambit perform bulk bitwise operations directly inside DRAM by exploiting internal analog behavior, achieving large gains for bitmap and bitset workloads at the cost of DRAM design changes and a fixed operation set~\cite{Seshadri2017Ambit}. Other near-memory accelerators in 3D-stacked or HBM-style memories similarly target specific primitives (\eg reductions, convolutions)~\cite{Mutlu2020ModernPrimerPM}. These approaches are powerful but typically \ci require specialized memory devices or non-standard DRAM interfaces, and \cii expose relatively rigid compute kernels rather than general, layout-centric transformations over high-dimensional objects. \tme is complementary: it assumes commodity DRAM and standard controllers, places logic on an \soc/\fpga fabric, and focuses on programmable cache line-level reorganization rather than adding arithmetic units inside memory arrays.
    
    \para{\fpga In-the-Middle and Near-Data Transformation.}
    Our design builds on the emerging idea of inserting reconfigurable logic directly in the memory path of \soc platforms. Programmable Logic In-the-Middle (PLIM)~\cite{Roozkhosh2020} and CAESAR~\cite{Roozkhosh2022CAESAR} show that an on-chip \fpga can sit between CPU clusters and main memory while remaining cache coherent, enabling fine-grained manipulation of memory traffic for resource management and QoS. Relational Memory (RME) leverages a similar \soc/\fpga setup to provide native, near-data projection: it transparently transforms row-wise base data into arbitrary column-groups on the fly, exposing both row and column access as if they were natively stored~\cite{Roozkhosh2023}. \tme generalizes this class of designs from relational projections to arbitrary dense tensor layouts. Unlike \plim/CAESAR, which targets control and routing of memory requests, or RME, which focuses on row/column projections for database tuples, \tme is explicitly architected as a tensor-centric cache line composer that supports a wide range of reshape, gather, and permutation patterns while maintaining a \pe-agnostic interface to the CPU. \section{Conclusion}
    \label{sec:conclusion}

    In this paper, we introduce the Tensor Memory Engine, a hardware–software co-designed approach that decouples computation from data layout (re)organization to improve locality in data-intensive applications. TME transparently reorganizes data on the CPU memory path without duplication, dynamically serving optimized cache lines to improve spatiotemporal locality. We implement TME on SoC/FPGA hardware, and our evaluation demonstrates that TME significantly enhances cache efficiency and performance without requiring application changes. By enabling dynamic data layouts and reducing the burden of manual tuning, TME provides a practical path toward more efficient memory systems for next-generation edge and data-intensive workloads. 
\balance
\bibliographystyle{plainurl}
\bibliography{references,library-manos,
additional-library-RW}

\end{document}